\documentclass[twocolumn,showpacs,preprintnumbers,amsmath,amssymb,prb,superscriptaddress]{revtex4-1}

\usepackage{graphicx}
\usepackage{color}
\usepackage[colorlinks=true,plainpages=false,linkcolor=blue,urlcolor=blue,citecolor=blue,pdfpagemode=UseNone,pdfstartview=FitBH]{hyperref}
\usepackage{array}

\def\SIO{$\rm Sr_2IrO_4$}
\def\jeff{$j_{\rm{eff}}$}
\def\SLIO{Sr$_{2-x}$La$_x$IrO$_4$}

\newcolumntype{P}[1]{>{\centering\arraybackslash}p{#1}}
\begin{document}

\title{Raman scattering study of vibrational and magnetic excitations in \SLIO}

\author{H.~Gretarsson}
\author{J.~Sauceda}
\author{N. H. Sung}
\affiliation{Max-Planck-Institut f\"{u}r Festk\"{o}rperforschung, Heisenbergstr. 1, D-70569 Stuttgart, Germany}
\author{M. H\"oppner}
\affiliation{Max-Planck-Institut f\"{u}r Festk\"{o}rperforschung, Heisenbergstr. 1, D-70569 Stuttgart, Germany}
\affiliation{QuantumWise A/S, Fruebjergvej 3, DK-2100 Copenhagen, Denmark}
\author{M. Minola}
\affiliation{Max-Planck-Institut f\"{u}r Festk\"{o}rperforschung, Heisenbergstr. 1, D-70569 Stuttgart, Germany}
\author{B. J.~Kim}
\affiliation{Max-Planck-Institut f\"{u}r Festk\"{o}rperforschung, Heisenbergstr. 1, D-70569 Stuttgart, Germany}
\affiliation{Department of Physics, Pohang University of Science and Technology, Pohang 790-784, Republic of Korea}
\author{B. Keimer}
\affiliation{Max-Planck-Institut f\"{u}r Festk\"{o}rperforschung, Heisenbergstr. 1, D-70569 Stuttgart, Germany}
\author{M. Le Tacon}
\affiliation{Karlsruher Institut f\"{u}r Technologie, Institut f\"{u}r Festk\"{o}rperphysik,  D-76344 Karlsruhe, Germany}

\date{\today}

\begin{abstract}
We have measured the doping and temperature dependence of lattice vibrations and magnetic excitations in the prototypical doped spin-orbit Mott insulator  \SLIO\ (x=0, 0.015, and 0.10). Our findings show that the pseudospin-lattice coupling -- responsible for the renormalization of several low energy phonon modes -- is preserved even when long-range magnetic order is suppressed by doping. In our most highly doped sample, the single magnon (Gamma-point) excitation disappears while the two-magnon mode softens and becomes heavily damped. Doping induced electron-phonon coupling is also observed in a higher energy phonon mode. We observe two different electron-phonon interaction channels, which provide evidence of the coexistence of fluctuating magnetic moments and mobile carriers in doped iridates.
\end{abstract}
 

\maketitle

\section{Introduction}

The notion that iridate perovskites from the Ruddlesden-Popper series are a spin-orbit analog of the high temperature superconducting cuprates has been widely discussed in recent years. \cite{FaWang_PRL2011,Watanabe_PRL_2013,Yang_PRB2014,Kee_PRL_2014} This view has largely been driven by experimental work on \SIO, that revealed a two-dimensional antiferromagnetic (AF) ground state\cite{BJKim_Science_2009} in which the ordered \jeff=1/2 pseudospin moments experience Heisenberg-like interactions.\cite{Jungho_PRL_2012} Akin to the cuprates, a transition from an insulating to a more metallic state occurs upon electron-doping \cite{Wilson_LaDoped_PRB_2015} followed by the appearance of Fermi arcs.\cite{Kim_Science2014,Baumberger_PRL_2015} At low temperatures ($\lesssim$ 80K), a $d$-wave gap in the electronic excitations \cite{Kim_dWave,Feng_dWave} was observed in angle-resolved photoemission spectroscopy (ARPES)  and scanning tunneling microscopy/spectroscopy (STM/STS) measurements on potassium decorated Sr$_{2}$IrO$_4$ surfaces --  results that are reminiscent of the pseudogap phase of the cuprates. These findings in surface doped Sr$_{2}$IrO$_4$ are a promising step towards the realization of high-temperature superconductivity, but they have only been partially reproduced in the bulk samples doped by chemical substitution. \cite{Baumberger_PRL_2015}

\begin{figure}[htb]
\includegraphics[width=1\columnwidth]{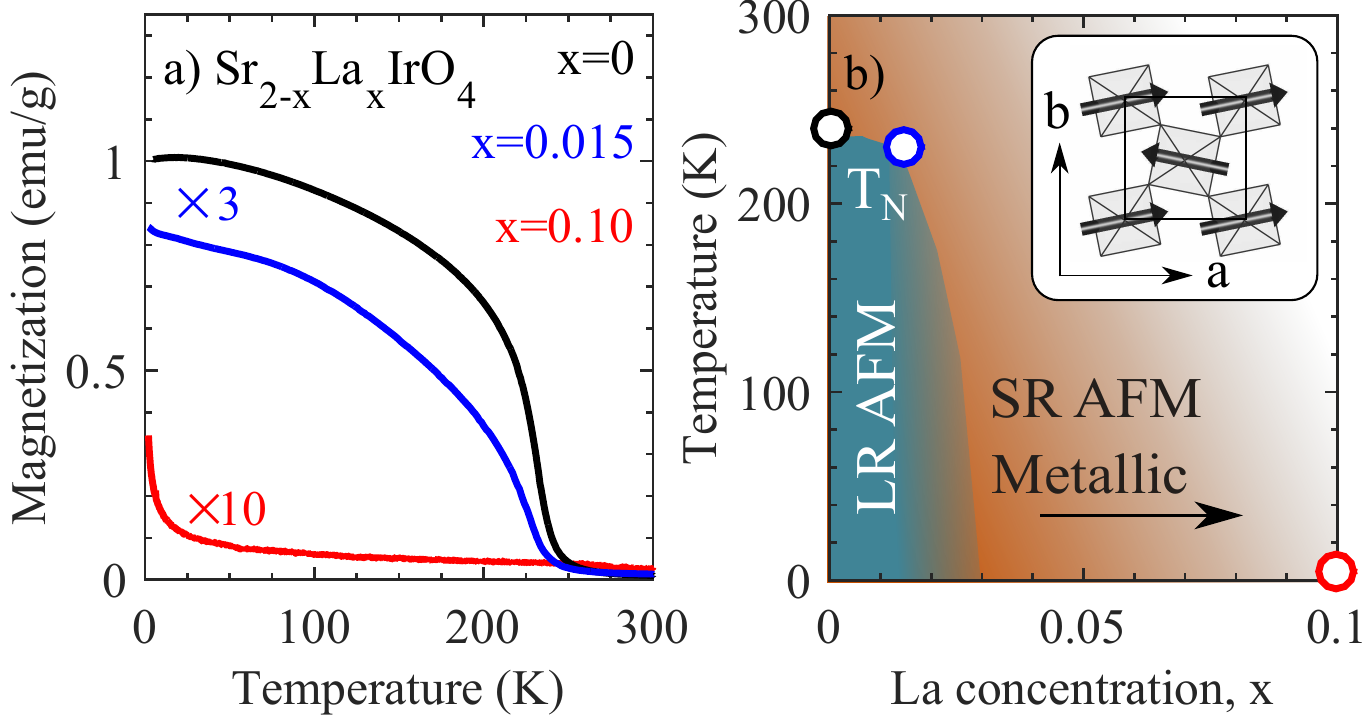}
\caption{\label{Phase} (a) Temperature dependent magnetization of Sr$_{2-x}$La$_x$IrO$_4$ in field cooled warming mode with applied field of H = 5 kOe perpendicular to the c axis. (b) Schematic phase diagram for Sr$_{2-x}$La$_x$IrO$_4$. Data was extracted from Ref. \onlinecite{Gretarsson_RIXS_LaDoping_2016}. Long range antiferromagnetic 
order is found in samples with $x=0$ and $0.015$ while only a short range antiferromagnetic order and a sizable Fermi surface is seen in $x=0.10$.   Inset shows the $ab$-plane of \SIO, arrows indicate magnetic order.}
\end{figure}

In bulk electron-doped \SLIO\, scattering experiments have demonstrated that, upon increasing the La-concentration, the magnetic long-range order (LRO) is rapidly replaced by a short-range order (SRO)  ($0.015\!\!<\!\!x_c\!\!\leq$0.04).   \cite{Wilson_LaDoped_PRB_2015,Gretarsson_RIXS_LaDoping_2016} This SRO phase persists up to at least x=0.10 and has been shown to host damped paramagnons,\cite{Gretarsson_RIXS_LaDoping_2016, Dean_arXiv_2016} again in striking similarity to the observations made in the cuprates.\cite{LeTacon_NatPhys_2011} Moreover, transport measurements reveal a weakly metallic state (x=0.04), and ARPES shows the formation of a Fermi surface at a doping level of x=0.10 (see phase diagram in Fig. \ref{Phase} (b)).\cite{Gretarsson_RIXS_LaDoping_2016} Although these findings arguably underscore the universality of the emergent low-energy physics in a doped Mott insulator, no consensus has been reached yet regarding the evolution of the charge gap in this doping range. For instance, recent ARPES experiments have shown a collapse of the Mott gap in a metallic sample along with the appearance of Fermi arcs\cite{Baumberger_PRL_2015}, whereas STM investigations indicate a phase-separated state, in which regions with a robust Mott gap are surrounded by metallic regions with a pseudogap.\cite{MPAllan_arXiv_2016}

The doping evolution of magnetic excitations and phonons can provide useful insights for this debate, either through their mutual coupling  \cite{Gretarsson_TwoMagnon_2015,Buchner_2016} or their possible interplay with mobile charge carriers. It has for instance been shown that pseudospin-lattice coupling exists in the paramagnetic state of \SIO, enabling one to monitor fluctuating moments through the lineshape of a selected phonon \cite{Gretarsson_TwoMagnon_2015} -- a task that is often challenging to carry out with other experimental techniques.\cite{Takagi_Elastic:PRL2012,Des_XY:PRB2015} 
Magnetic excitations and phonons provide a unique view into how the spin-orbit entangled ground state in \SIO\ responds to doping. Lowering of the crystal symmetry, either through a structural transition or charge order, can result in additional phonon modes, while variations in magnetic exchange interactions are reflected directly in the magnetic excitations.
Whereas progress has been made regarding the doping evolution of the magnetic structure and dynamics in doped iridates, \cite{Gretarsson_RIXS_LaDoping_2016,Dean_arXiv_2016,MDean_RIXS:arXiv2016,Calder_RIXS:PRB2016} only little is known about the lattice dynamics of these materials.\cite{Sr214_Ru_Raman}

In this paper we address this issue and present a comprehensive Raman scattering study of the doping dependence of the lattice dynamics and magnetic excitations in electron-doped Sr$_{2-x}$La$_x$IrO$_4$ ($x=0$, $0.015$, and $0.10$). As we enter the SRO metallic phase ($x=0.10$) selected phonon modes show a spectacular evolution of their $low$ $temperature$ lineshapes, reflecting the stronger fluctuations of pseudospins, while the changes at $high$ $temperatures$, where the pseudospins remain dynamic at all doping levels,   are negligible. Additionally, we observe the emergence of a novel Fano line shape asymmetry for a higher-energy phonon mode. This demonstrates the existence of a continuum of excitations that were not present in the parent compound, and that we assess to the electron-hole continuum of the metallic state. From the spin dynamics point of view, the single magnon (Gamma-point) peak originally reported in ref.~\onlinecite{Cooper_PRB_Raman_2016} rapidly disappears, whereas the two-magnon excitation survives, although heavily damped and softer.

\section{Experimental Details}

\subsection{Samples}

\SLIO\ crystals ($x=0$, $0.015$, and $0.10$) were grown using a flux method \cite{Nakheon_Growth_2016} and  the La concentrations were checked via electron probe micro-analysis. In Figure \ref{Phase}(a) we show the temperature dependent magnetization of Sr$_{2-x}$La$_x$IrO$_4$  in an external magnetic field of H~=~5~kOe perpendicular to the crystallographic $c$-axis. In crystals with $x=0$ and 0.015, the onset of a weak ferromagnetic response due to the formation of the canted antiferromagnetic order \cite{BJKim_Science_2009} is seen around $T_N =$ 240 K and 230 K, respectively. This ferromagnetic response is absent in the $x=0.10$ sample, indicating a doping-induced transition into a paramagnetic state.  ARPES measurements on a $x=0.10$ sample shows a clear Fermi surface, confirming that our most highly doped sample is in the metallic phase. These findings are summarized in the phase diagram in Fig. \ref{Phase}(b).

\begin{figure}[htb]
\includegraphics[trim=0cm 0cm 0cm 0cm, clip=true, width=.85\columnwidth]{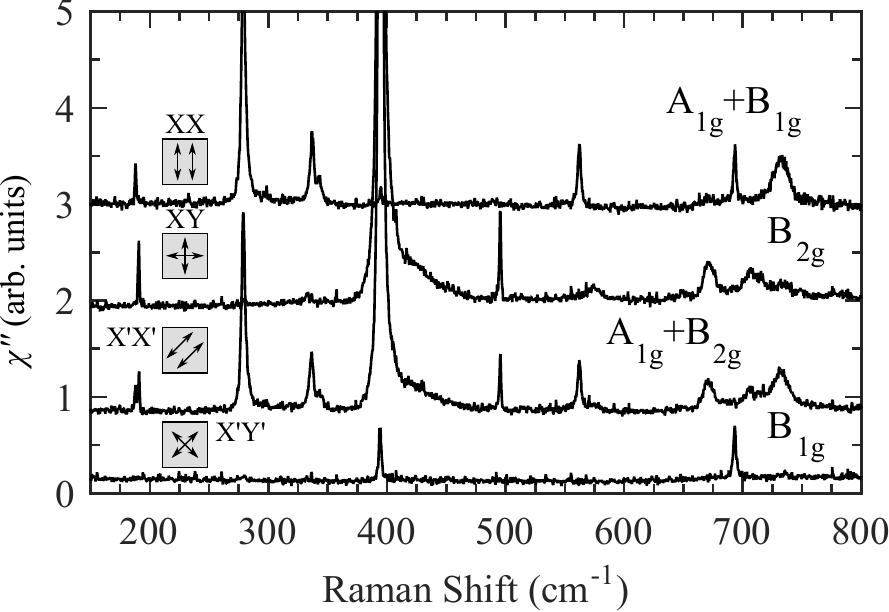}
\caption{\label{Selection} Raman spectra of \SIO\ taken at $T = 50$ K using four different in-plane polarization. The arrows in the insets indicate the electric field vectors of the incident and the scattered light with respect to the unit cell of \SIO.  XX-channel probes the $A_{1g}+ B_{1g}$ modes, XY the $B_{2g}$, X$^\prime$X$^\prime$ the $A_{1g} + B_{2g}$, and X$^\prime$Y$^\prime$ the $B_{1g}$ . }
\end{figure}

\begin{table*}

\begin{tabular*}{\textwidth}{@{\extracolsep{\fill} } P{1.5cm} | P{1.5cm} | P{1.35cm} P{1.35cm} P{1.35cm} P{1.35cm} | P{2cm} | P{3cm} | P{3cm}  }
\hline \hline
 & & \multicolumn{4}{c|}{Polarization}  &   &  Comment  & Comment  \\
Calc.  & Exp.  & XX &  XY & X$^\prime$X$^\prime$ & X$^\prime$Y$^\prime$  & Assignment  & Parent & Doped \\
(cm$^{-1}$) &(cm$^{-1}$) & $A_{1g}+B_{1g}$ & $B_{2g}$ & $A_{1g}+B_{2g}$ & $B_{1g}$    & of modes & x=0  & x=0.10  \\
\hline
110 &        &  &  &  & & $B_{1g}$ & Not detected & Not detected \\
135 &        &  &  &  & & $B_{2g}$ & -- & --\\
168 &        &  &  &  & & $B_{1g}$ & -- & --\\
188 & 188       & \checkmark &  & \checkmark & & $A_{1g}$ & Fano $T>T_N$ & Fano at all $T$  \\
171 & 191          & & \checkmark &  \checkmark & & $B_{2g}$ & -- & --  \\
326 & 278           & \checkmark &  & \checkmark & & $A_{1g}$ & Very broad at high $T$ & Broad at all $T$ \\
& 337           & \checkmark &  & \checkmark & & $A_{1g}$ & Only low $T$ & Not detected \\
& 342           & \checkmark &  & \checkmark & & $A_{1g}$ & -- & -- \\
391 &        &  &  &  & & $B_{1g}$ & Not detected & Not detected\\
410 & 395          & & \checkmark &  \checkmark & & $B_{2g}$ & Very intense & Weaker + Fano \\
457 & 495          & & \checkmark &  \checkmark & & $B_{2g}$ & Very weak at high $T$ & Very weak \\
532 & 562          & \checkmark &  & \checkmark & & $A_{1g}$ & Detected & Detected\\
546 &        &  &  & &   & $B_{1g}$ & Not detected & Not detected\\
660 & 693       & \checkmark &  & &  \checkmark & $B_{1g}$ & Weak at high $T$ & Not detected \\

\hline
\hline
\end{tabular*}
\caption{Comparison between the Raman active modes obtained from first-principles lattice
dynamics calculations and the experimental phonon modes in Fig.\ref{Selection}.}
\label{Table_Phonon}
\end{table*}

\subsection{Raman Measurements}

The Raman scattering experiments were performed using a JobinYvon LabRam HR800 spectrometer and the 632.8 nm excitation line of a HeNe laser.  Two configurations were used, low-resolution (600 grooves/mm grating yielding energy resolution $\sim$ 5 cm$^{-1}$) and high-resolution (1800 grooves/mm  $\sim$ 1.5 cm$^{-1}$). Due to the weak intensity of the lowest-energy phonon modes in \SIO\ the  power of the laser was kept at 2 mW, for all other measurements a power of 0.7 mW was used. The diameter of the beam was $\sim 10$ $\mu$m. Measurements were corrected for laser heating (0.7 mW: $\Delta T\sim$ 20 K, 2 mW: $\Delta T\sim$ 40 K). The heating was determined experimentally by measuring the ratio of the anti-Stokes to Stokes intensities of the single-magnon excitation at the lowest temperature of the cryostat. The phonon and two-magnon spectra were Bose corrected ($\chi^{\prime\prime}$). This was done by dividing the total measured intensities by the Bose-Einstein factor, $(1+n(\omega))$, where $n(\omega)$ represents the Bose distribution.  Additionally, a correction accounting for the spectrometer response was applied.  The samples were placed in a He-flow cryostat, and the measurements were carried out in backscattering geometry with the light propagating along the crystalline c-axis, while the polarization of the incident and scattered light was varied within the $ab$-plane. Because of rotation of the IrO$_6$ octrahedra around the c-axis in Sr$_2$IrO$_4$, a- and b-axes are taken along the Ir-Ir next nearest-neighbor direction (see inset Fig. \ref{Phase} b)  (space group $I4_1/acd$ $a=b\approx 5.5 {\rm \AA}$, $c\approx 26  {\rm \AA}$).\cite{Sr214_Structure}  The Raman spectrum in the XX-channel probes the $A_{1g} + B_{1g}$ symmetry modes while the XY-channel probes the $B_{2g}$ symmetry modes. A schematic picture of the incident and scattered light polarization in these two experimental configurations can be seen in inset in Fig. \ref{Selection}.  To obtain a clean surface before the measurements, the samples were cleaved $ex$ $situ$.

\subsection{Selection Rules}

In Fig. \ref{Selection} we plot the Raman spectra of \SIO\  at $T~=~50$ K  in $A_{1g}+B_{1g}$, $B_{2g}$, $A_{1g}+B_{2g}$, and $B_{1g}$ scattering geometries. Based on first-principles lattice dynamics  calculations (details are given in ref. \onlinecite{Propper_Optics:PRB2016}) we expect $3A_{1g}$, $5B_{1g}$, and $4B_{2g}$ modes in these geometries. The observed and calculated phonon modes and their symmetries are listed in Table \ref{Table_Phonon}. The broad modes around 700 cm$^{-1}$ are not included and originate most likely from two-phonon scattering. We are able to observe $5A_{1g}$, $1B_{1g}$, and $3B_{2g}$ modes. No phonons were observed below 150 cm$^{-1}$ (not shown).  Compared to earlier studies on \SIO\ \cite{Sr214_Raman} we see  two new $A_{1g}$ phonon modes at 337 cm$^{-1}$ and 342 cm$^{-1}$ and a new $B_{2g}$ mode at 191 cm$^{-1}$. The two new $A_{1g}$ phonon modes develop however only at low-temperatures and are not predicted by the calculations. These modes will be addressed later in the paper. We assign the mode at 395 cm$^{-1}$ in the $B_{1g}$ channel to leaking of the intense $B_{2g}$ mode. The agreement between calculated and measured phonon energies is reasonable although one out of the four $B_{2g}$ and most of the  $B_{1g}$ phonon modes have not been detected. The energy discrepancy between the calculated and measured phonon frequencies is likely due to the metallic nature of the ground state in this lattice dynamics calculation. The phonon energies are therefore affected by charge screening, which is not present in the insulating \SIO.

\section{Experimental Results} 

\begin{figure*}[htb]
\includegraphics[ trim=0cm 0.75cm 0cm 0cm, clip=true, width=2\columnwidth]{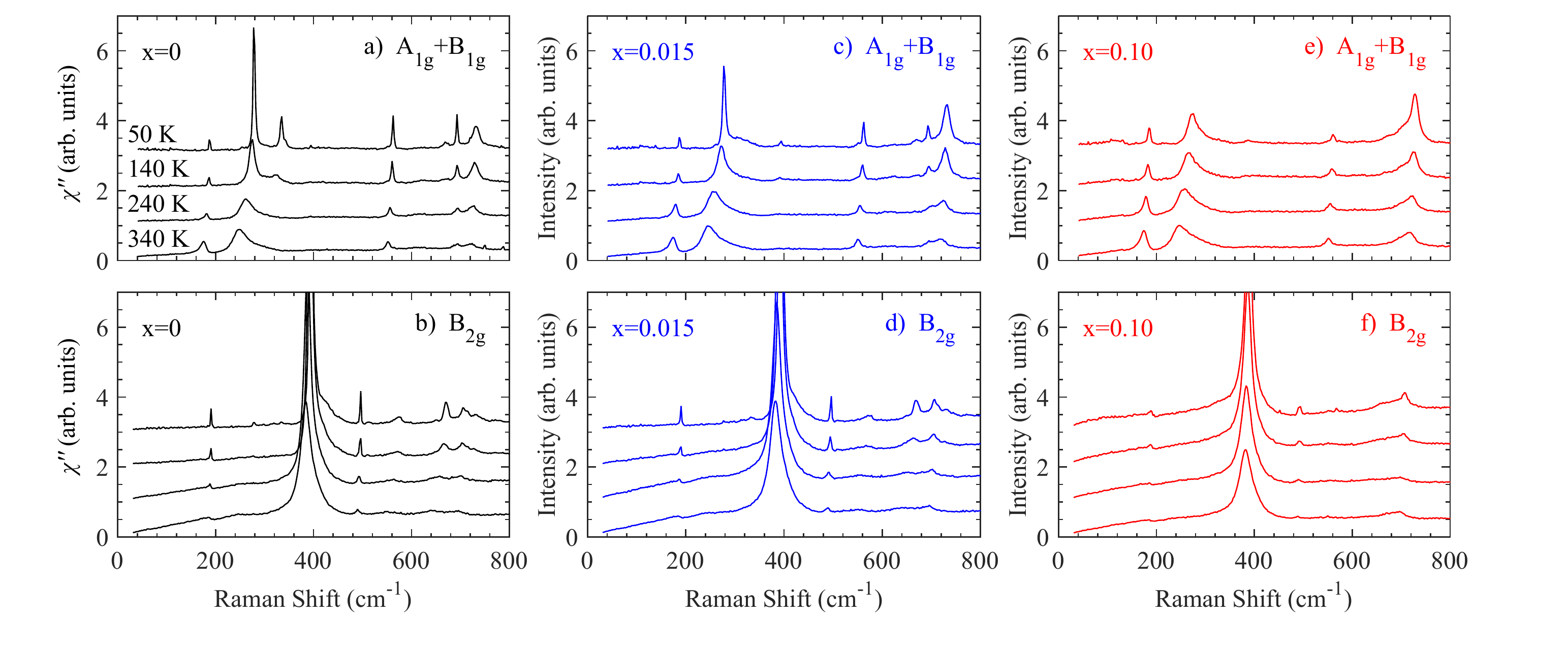}
\caption{\label{Allphonon} Doping and temperature dependence of the $A_{1g}+B_{1g}$ (top: a, c, and e)  and the $B_{2g}$  (bottom: b, d, and f) phonon modes in \SLIO.}
\end{figure*}

We have performed temperature-dependent Raman scattering measurements on single crystals of \SLIO\  ($x=0$, $0.015$, and $0.10$). Our main results can be seen in Fig. \ref{Allphonon}. First, we observe no doping induced Raman modes, which indicates that the crystal structure of the parent compound remains intact upon doping, in agreement with prior x-ray studies.\cite{Wilson_LaDoped_PRB_2015} Second, several phonons modes experience a strongly doping dependent lineshape renormalization as  the La-content is increased. In some cases, an asymmetric  ``Fano'' lineshape that characterizes  coupling of the phonon mode to a continuum of excitations, is induced by the La doping. These modes will be more specifically discussed in the following. To quantify our observation, we have fitted these phonon modes with a Fano profile \cite{Fano:PR1961} described by the formula:

\begin{equation}
 I(\omega)=I_0(q+\epsilon)^2/(1+\epsilon^2)+I_{back.}(\omega),
 \label{Fanoformula}
\end{equation}

\begin{figure}[htb]
\includegraphics[ trim=0.25cm 0cm 0cm 0cm, clip=true, width=0.85\columnwidth]{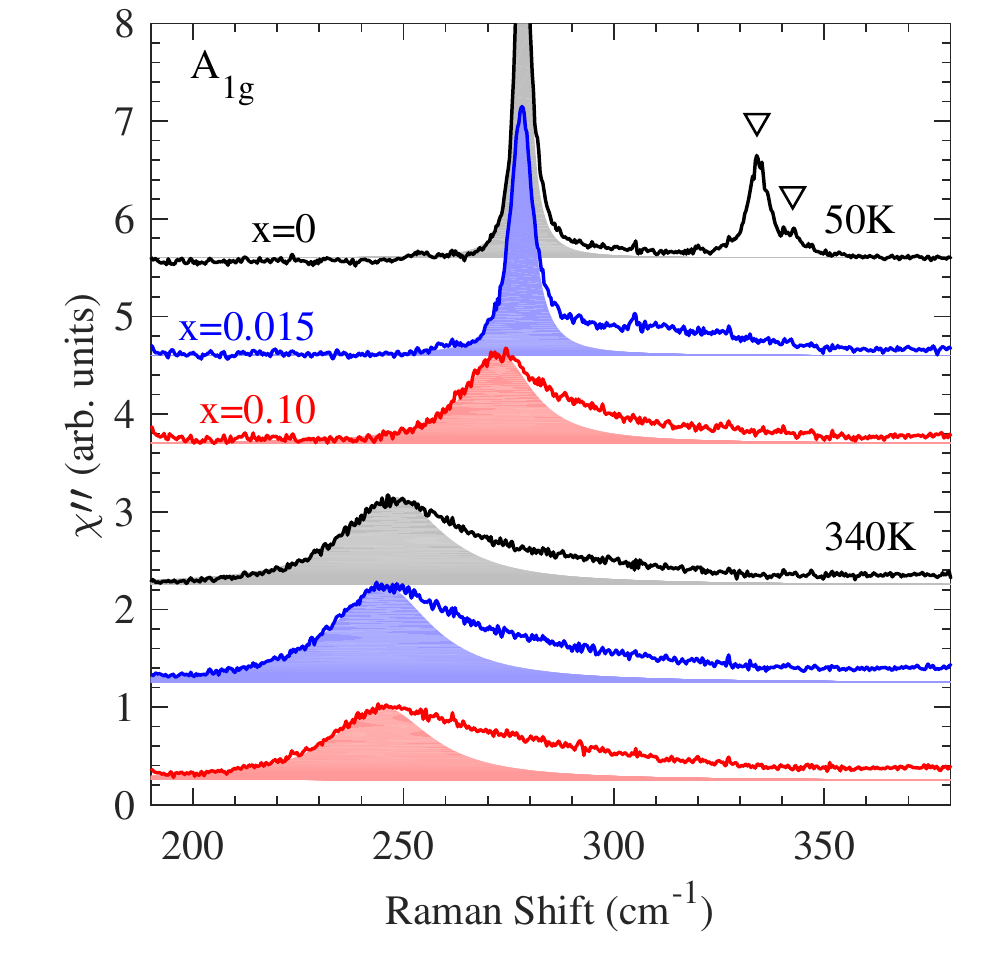}
\caption{\label{A1g_newPh} Detailed view of the doping and temperature dependence of the 278 cm$^{-1}$ $A_{1g}$ mode, included are the two new modes at 337 cm$^{-1}$ and 342 cm$^{-1}$ (triangle). The filled Lorentzian function represents the contribution from the 278 cm$^{-1}$ mode. }
\end{figure}

\noindent where $\epsilon=(\omega-\omega_0)/\Gamma$, $\omega_0$ is the bare phonon frequency, $\Gamma$ is the linewidth, $q$ is the asymmetry parameter, and $I_{back.}$ is a linear background. The inverse of the asymmetry parameter,  $1/|q|$, correlates with the strength of the coupling between the discrete phonon mode and the underlying charge and/or spin continuum.

\subsubsection{A$_{1g}$ phonon modes}

In Fig. \ref{Allphonon}(a) the temperature dependence of the spectra measured in the $A_{1g}$+$B_{1g}$ symmetry in the parent compound are plotted. At low temperatures, the spectrum is dominated by the $A_{1g}$ mode at 278 cm$^{-1}$ , which has been assigned to the bending of the in-plane Ir-O-Ir bonds.\cite{Sr214_Raman} In Fig. \ref{A1g_newPh} a detailed view of this mode can be seen along with a filled Lorentzian function representing its contribution. The  278 cm$^{-1}$ phonon shows a large softening ($\sim$ 10$\%$ from $T=50$ K to $340$ K)\cite{Sr214_Raman} and has a broad lineshape at high-temperatures. Opposite to the case of  Sr$_2$Ir$_{1-x}$Ru$_x$O$_4$ \cite{Sr214_Ru_Raman} and Sr$_2$IrO$_{4-\delta}$, \cite{Nakheon_Growth_2016} we see no evidence of doubling of this mode upon doping. Instead, this mode softens ($\sim$ 2$\%$) and its broad lineshape is retained at all temperatures in the highest doped sample. At a slightly higher energy we observe the two weaker $A_{1g}$ modes, at 337 cm$^{-1}$ and 342 cm$^{-1}$ (triangle symbol in Fig. \ref{A1g_newPh} ). These modes depend strongly on temperature and doping. In particular, they broaden significantly at high temperatures and appear to merge into the high-energy tail of the 278 cm$^{-1}$ mode. In the doped material, they do not exhibit the strong temperature dependence encountered in the parent compound (see Fig. \ref{Allphonon}(c)) and instead form a broad hump (see x=0.015 and $T = 50$ K in Fig. \ref{A1g_newPh}) . Their weak intensity however prevented us from studying them accurately enough.

\begin{figure}[htb]
\includegraphics[ trim=0cm 0cm 0cm 0cm, clip=true,width=0.95\columnwidth]{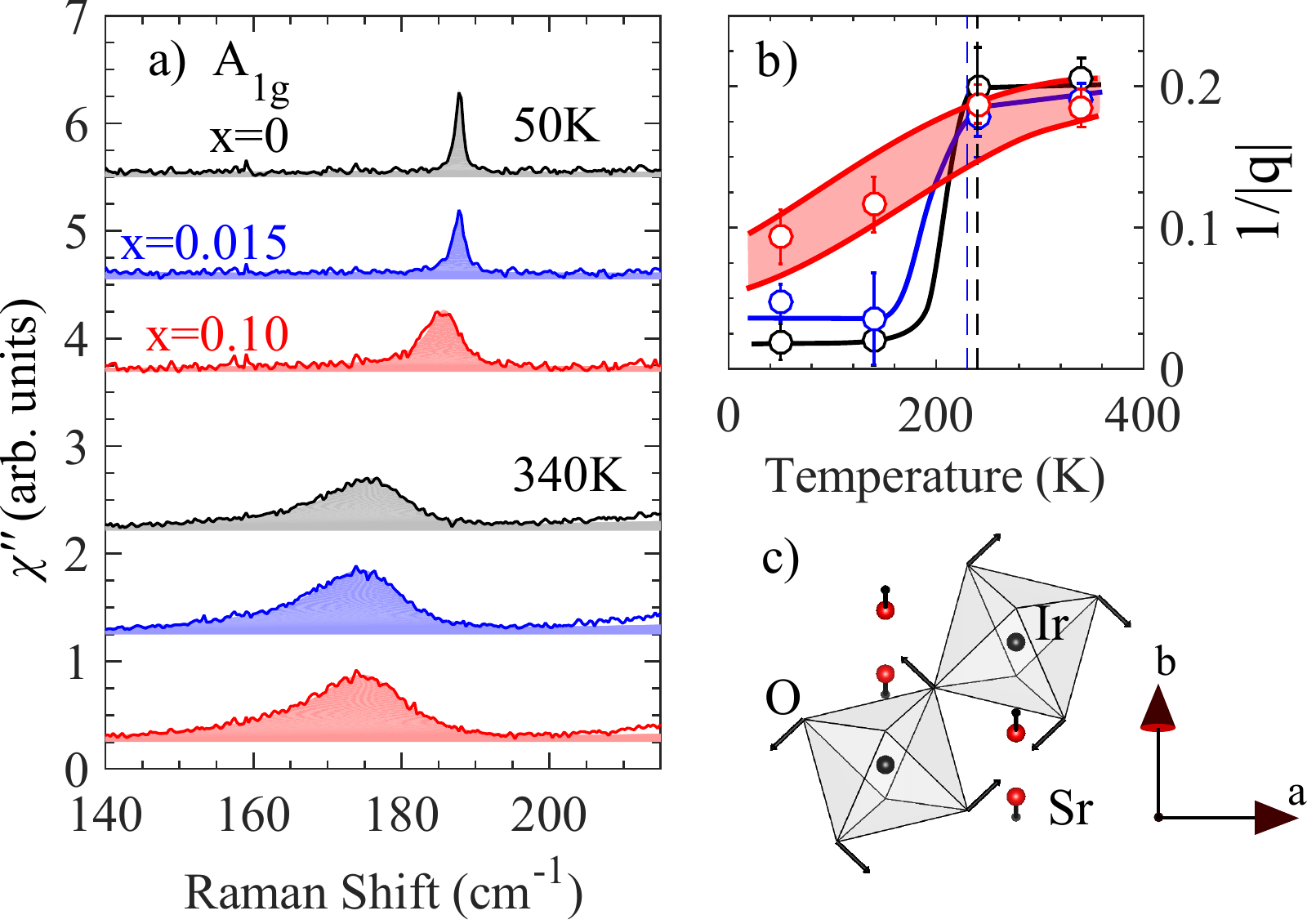}
\caption{\label{A1g_Phononfit}  Doping dependence of the (a) 188 cm$^{-1}$ $A_{1g}$ phonon mode taken at high- and low-temperatures. Spectrum have been shifted for clarity. The filled area is a result of a fit (see text). (b) Temperature dependence of the inverse Fano asymmetry parameter, $1/|q|$. The dashed vertical lines represent $T_N$  for $x=0$ and $0.015$ samples. Solid lines are a guide to the eye. (c) Calculated eigenvector of the 188 cm$^{-1}$ $A_{1g}$ twisting mode.}
\end{figure}

At low energies we observe the $A_{1g}$ 188 cm$^{-1}$ phonon mode in all samples, albeit with different lineshapes. In the parent compound this mode has been shown to interact with a continuum of pseudospin fluctuations.\cite{Gretarsson_TwoMagnon_2015} In Fig. \ref{A1g_Phononfit}(a) the doping dependence of this mode is plotted for $T = 50$ K and $T = 340$ K along with the calculated eigenvector (panel (c)). Like the 278 cm$^{-1}$ $A_{1g}$ phonon, this mode modulates the important $\rm{Ir-O-Ir}$ in-plane bond, which is intimately tied to the magnetic order (see inset in Fig. \ref{Phase}(b)). In the samples showing LRO ($x=0$ and $0.015$) this low-energy $A_{1g}$ mode is sharp and symmetric at $T = 50$ K while it is asymmetric in the sample showing SRO ($x=0.10$). By fitting the phonons (shaded area in (a)) using Eq. \ref{Fanoformula} we can extract the inverse of the asymmetry parameter, $1/|q|$, as a function of temperature and doping.  In Fig. \ref{A1g_Phononfit}(b) we plot our fitting results. While the results on the $x=0$ and $0.015$ samples show that $1/|q|$ is suppressed in the magnetically ordered state, only a gradual reduction is observed in the $x=0.10$ sample, in which the asymmetry remains visible down to the lowest temperatures. Remarkably, although a metallic state has been reached in the $x=0.10$ sample, we observe very little difference in the line shape of the phonons at $T = 340$ K. This observation further strengthens the idea that the underlying continuum, which couples to the 188 cm$^{-1}$ $A_{1g}$ phonon and is gapped out for $T< T_N$ for samples with magnetic LRO, does not change significantly with doping and is still dominated by pseudospin fluctuations.\cite{Gretarsson_TwoMagnon_2015}

\subsubsection{B$_{2g}$ phonon modes }

\begin{figure}[htb]
\includegraphics[ trim=0cm 0cm 0cm 0cm, clip=true,width=1\columnwidth]{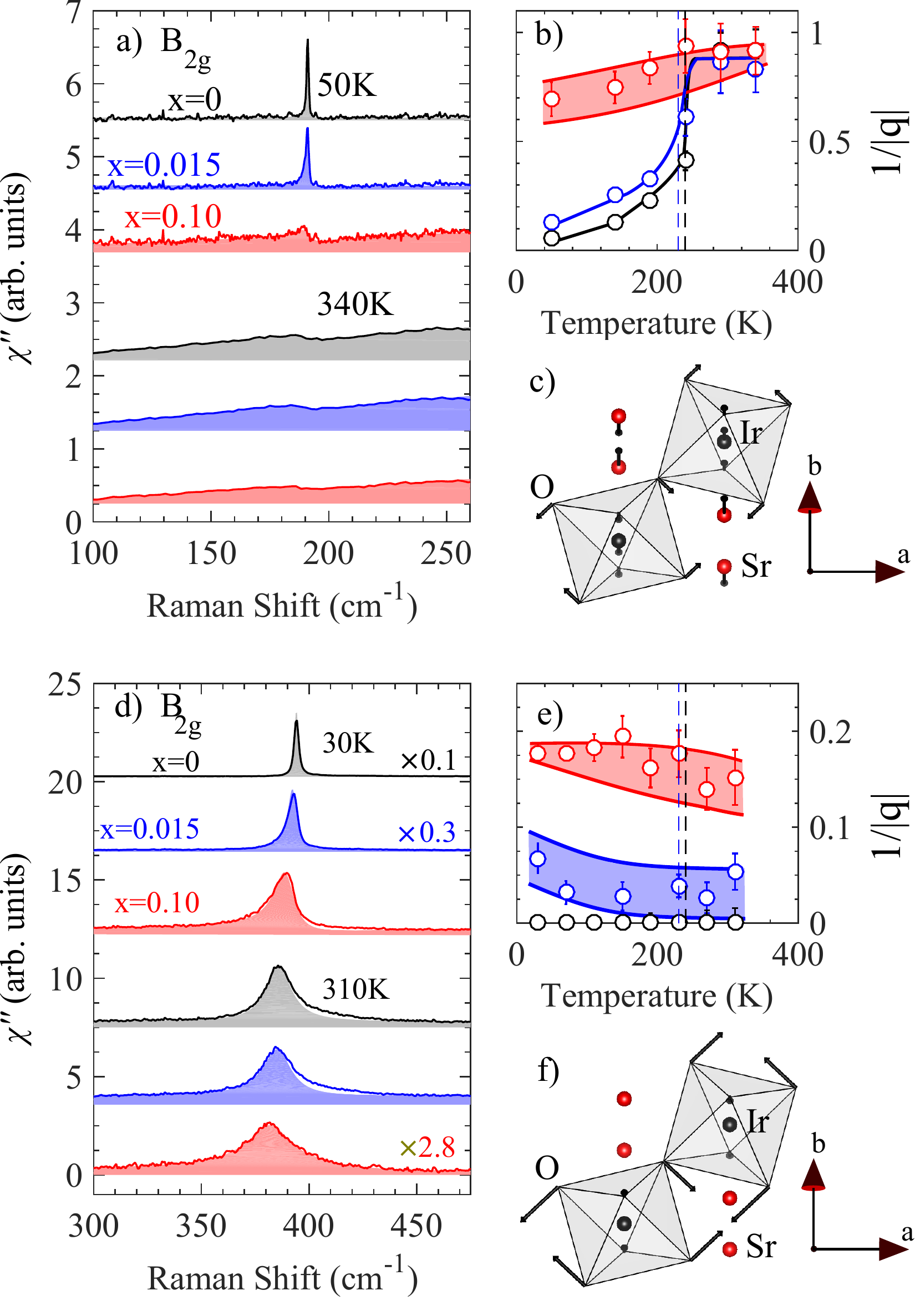}
\caption{\label{B2g_Phononfit}  Doping dependence of the (a) 191 cm$^{-1}$ and (d) 395 cm$^{-1}$ $B_{2g}$ phonon modes taken at high- and low-temperatures. Spectrum have been shifted and scaled in intensity (where indicated) for clarity. The filled area is a result of a fit (see text). (b,e) Temperature dependence of their inverse Fano asymmetry parameter, $1/|q|$. The dashed vertical lines represent $T_N$  for $x=0$ and $0.015$ samples. Solid lines are a guide to the eye. (c,f)  The calculated eigenvectors.}
\end{figure}

The B$_{2g}$ Raman spectra are shown in Fig. \ref{Allphonon}(b,d,f). For all samples the  spectra are dominated by an intense phonon at 395  cm$^{-1}$ associated with in-plane stretching of the IrO$_6$ octahedra \cite{Gretarsson_TwoMagnon_2015} (see Fig. \ref{B2g_Phononfit}(f) for the calculated eigenvector).  At lower energies the new 191 cm$^{-1}$  $B_{2g}$ phonon mode can be seen. Both modes react strongly to temperature and doping.

In Fig. \ref{B2g_Phononfit} we compare the (a) 191 cm$^{-1}$ and (d) 395 cm$^{-1}$   $B_{2g}$ phonon modes in the $x=0$, $0.015$, and $0.10$ samples as a function of temperature. The shaded areas are results of fitting the spectra to Eq. \ref{Fanoformula}. We note that for the 395 cm$^{-1}$ mode additional spectral weight on the higher-energy side is visible in the  $x=0$ and $0.015$ samples  (see unfilled fitted area). This spectral weight is also present at low temperatures and cannot be fitted properly using a Fano line shape. At x=0.10, it is however indistinguishable from the broad phonon mode. The eigenvectors of both of these modes involves in-plane stretching of the O-atoms (c,f), but only the 191 cm$^{-1}$ mode (c) has $c$-axis modulation of Sr- and Ir-atoms. By looking at the result of our fit  in Fig. \ref{B2g_Phononfit}(b,e) it is evident that these two $B_{2g}$ modes behave differently. While the 191 cm$^{-1}$ mode shows results reminiscent of the 188 cm$^{-1}$ $A_{1g}$  mode, the 395 cm$^{-1}$ mode acquires an anisotropic line shape upon doping. 

As previously reported, the 395 cm$^{-1}$ mode of the parent compound has indeed a Lorenzian lineshape and follows a regular  anharmonic decay.\cite{Gretarsson_TwoMagnon_2015} However, upon electron doping at low temperature this mode exhibits a marked asymmetrical lineshape. Interestingly, and unlike the effect observed on the 188 cm$^{-1}$ $A_{1g}$ and 191 cm$^{-1}$ $B_{2g}$ modes, this doping-induced lineshape change shares no similarities with the thermal effect in the $x=0$ sample and the asymmetric line shape shows very little temperature dependence. This strongly suggests a different physical origin for this behavior. Before discussing this in the next section, we now describe the effects of doping on the magnetic excitations observed by Raman scattering.

\begin{figure}[htb]
\includegraphics[ trim=1cm 0.5cm 17.5cm 0cm, clip=true,width=1\columnwidth]{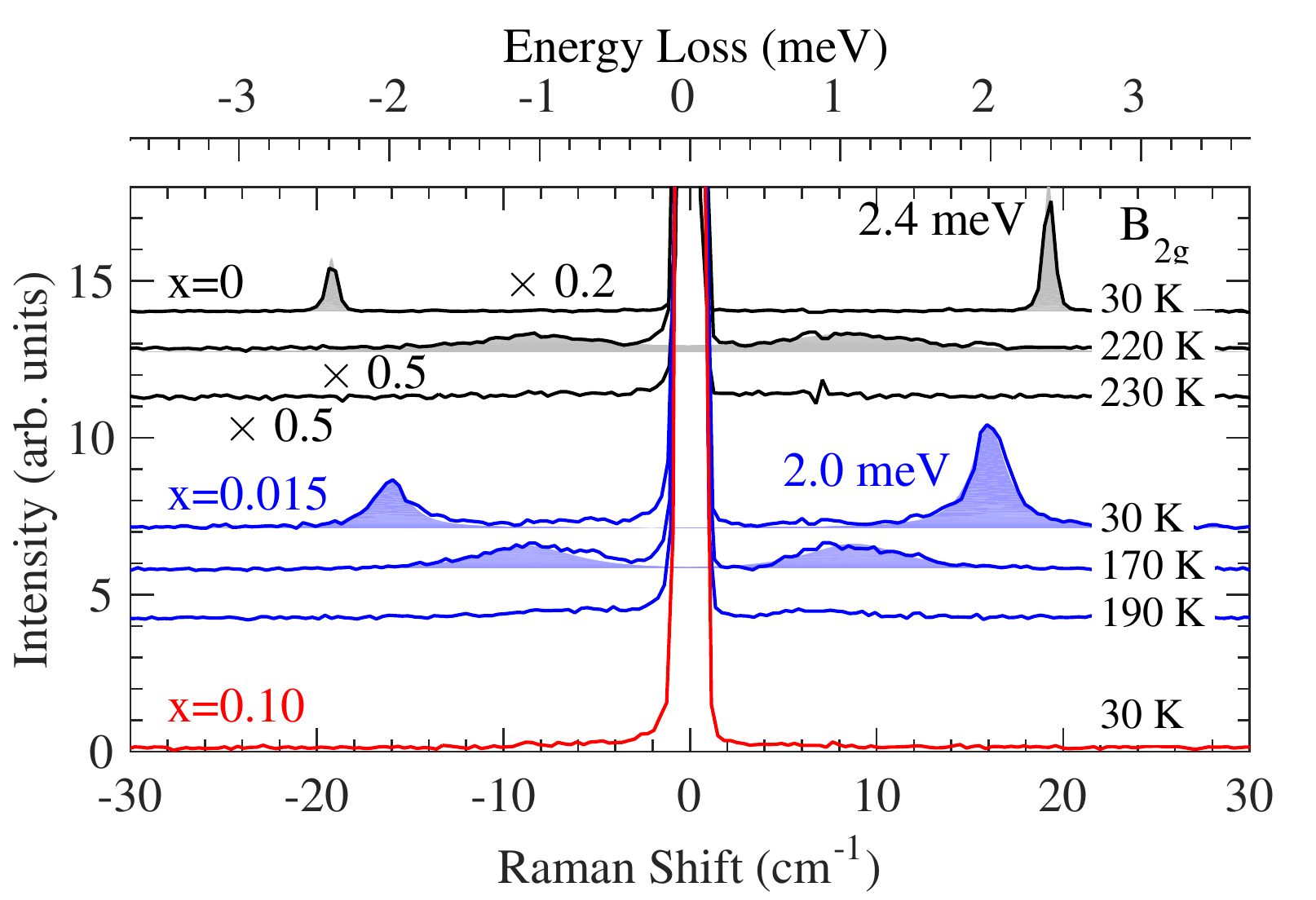}
\caption{\label{Magnon}  Low energy Raman spectra  of the $B_{2g}$ single-magnon scattering in \SLIO\  at  selected temperatures. Both the Stokes and anti-Stokes part are shown. Spectrum have been shifted and scaled in intensity (where indicated)  for clarity. The filled Lorentzians are a results of a fit (see text).  
}
\end{figure}

\subsubsection{Single-magnon excitations}

The phonon anomalies observed across the magnetic transitions  (Fig. \ref{A1g_Phononfit}(b) and \ref{B2g_Phononfit}(b)) indicate a strong interplay between the lattice and magnetic degrees of freedom. In the $B_{2g}$ channel, zone-center single magnon excitations have been previously reported in  \SIO.\cite{Cooper_PRB_Raman_2016}

In Fig. \ref{Magnon} we plot the low energy Raman spectra in the  $B_{2g}$ channel, both the energy gain (anti-Stokes) and energy loss sides (Stokes) at selected temperatures. For the parent compound we see a single resolution-limited peak  which is also visible on the energy gain side. To extract the peak position of the $x=0$ magnon we have fitted the spectrum with two resolution-limited Lorentzian functions, constrained by the principle of detailed balance  (temperature was fixed at $T = 30$ K). This leaves us with two fitting parameters: the amplitude and peak position ($E_m$).  In Fig. \ref{Magnon} we show our fit result for $x=0$ as a shaded gray area where our results give $E_m=19.2$ cm$^{-1}$ ($\sim$ 2.4 meV).  The energy position is in good agreement with a previous report on \SIO.\cite{Cooper_PRB_Raman_2016} For the $x=0.015$ sample the magnon is significantly weaker and broader. By using the same fit as for $x = 0$ (width is also fitted) we get $E_m=17$ cm$^{-1}$ ($\sim$ 2.1 meV). We note that the very weak hump around 7 cm$^{-1}$ in the $T= 190$ K spectrum of the $x=0.015$ sample is also there at $T = 210$ K, with no change in intensity or shape, indicating that the hump is not coming from magnetic scattering.
In the $x=0.10$ sample we find no magnetic signal at low energies. 

In Fig. \ref{Magnonfit} (a) the integrated intensity of the single magnon as a function of temperature has been plotted. As is seen in the raw data for the $x=0$ and $0.015$ samples in Fig. \ref{Magnon}, no evidence of spin excitations can be observed at $T = 230$ K and $190$ K, respectively. The onset temperature of the magnon in $x=0$ agrees well with the magnetic ordering temperature extracted from the susceptibility data in Fig. \ref{Phase}(a) but it is lower in the $x=0.015$ sample.  Having carefully ruled out the effects of laser heating, we conclude that the onset temperature of the single-magnon is lower than $T_N$ in our doped sample.

\subsubsection{Two-magnon excitations}
\begin{figure}[htb]
\includegraphics[ trim=16.5cm 0.5cm 1cm 0cm, clip=true,width=1\columnwidth]{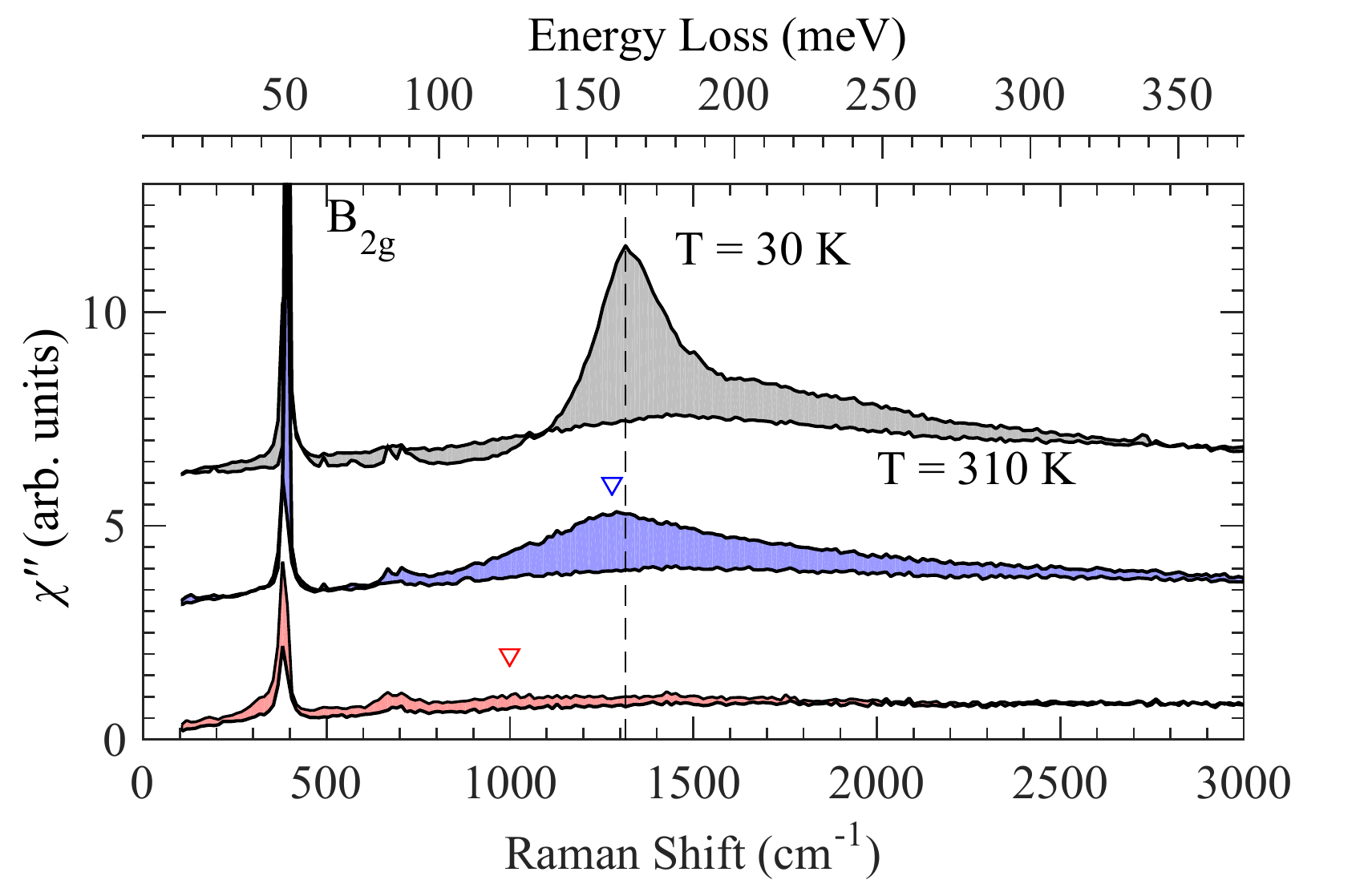}
\caption{\label{Bimagnon} High-energy Raman spectra of the $B_{2g}$ modes showing the two-magnon scattering in \SLIO\ as a function of doping. Spectrum have been shifted for clarity. For each doping level the spectrum at $T = 30$ K and  $T = 310$ K is shown. The filled area emphasizes the change in spectral weight upon cooling. }
\end{figure}

The absence of a magnetic signal at low energies in the $x=0.10$ sample in Fig. \ref{Magnon} is expected since the long range magnetic order is lost in this sample. The two-magnon scattering, on the other hand, weights the two-magnon density of states by a structure factor that mostly probes the excitations close to the Brillouin zone boundary, and is therefore an incisive probe of magnetism on shorter length scales. 
Like the single magnon, the two-magnon scattering is observable in the  $B_{2g}$  channel,\cite{Gretarsson_TwoMagnon_2015} albeit at a much higher energy loss that reflects the strength of the magnetic exchange. In Fig. \ref{Bimagnon} we plot the $B_{2g}$ Raman spectra of \SLIO\ between 100 and 3000 cm$^{-1}$ for $T = 30$ K and  $T = 310$ K. At $T = 30$ K the two-magnon scattering in \SIO\ is centered around 1300 cm$^{-1}$ ($\sim$ 160 meV)  and is also visible at $T = 310$ K. In the broken-bond model,\cite{Loudon_Fleury} the creation of two-magnons in a 2D Heisenberg S=1/2 system results in an energy cost of $E_{2M}=3J$, which is reduced to $\sim2.76J$ by quantum fluctuations.\cite{Weber_TwoMagnon:PRB1989} For $E_{2M}=160$ meV, this yields $J=60$ meV in excellent agreement with RIXS results.\cite{Jungho_PRL_2012}

\begin{figure}[htb]
\includegraphics[width=1\columnwidth]{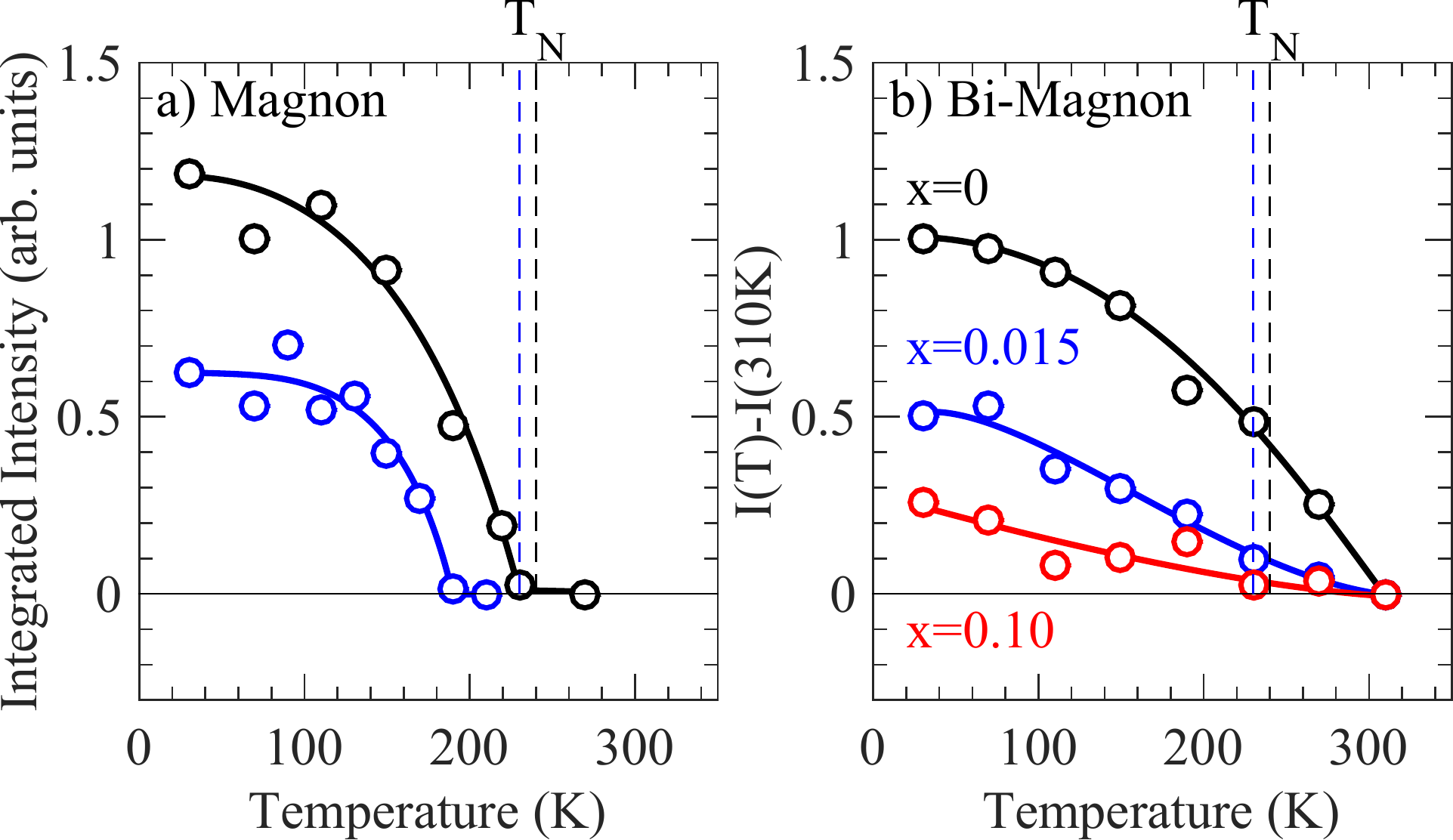}
\caption{\label{Magnonfit} (a) Temperature dependence of the integrated intensity of the  (a) single magnon excitation in  \SLIO\  ($x=0$ and $x=0.015$) and (b) two-magnon scattering ($x=0$, $x=0.015$ and $x=0.10$). Solid lines are a guide to the eye. In (a) the background at high temperature has been subtracted, while in (b), where the two-magnon persists to high temperature, the change in the integrated intensity is plotted with respect to $T  = 310$ K. Dashed vertical lines represent Neel temperature, determined from susceptibility measurements on the $x=0$ and $x=0.015$ samples.   }
\end{figure}

Upon slight electron doping, $x=0.015$,  the sharp and asymmetric two-magnon scattering softens visibly by about 40 cm$^{-1}$ (see blue triangle pointing to the peak position) before becoming significantly weaker and softer by about 300 cm$^{-1}$ (center of mass $\sim 1000$ cm$^{-1}$)  at $x=0.10$. The small softening in the $x=0.015$ sample translates to $\sim3\%$ reduction in the nearest-neighbor coupling $J$, while translating to $\sim20\%$ reduction in the $x=0.10$  sample, values that are consistent with recent RIXS results on \SLIO.\cite{Gretarsson_RIXS_LaDoping_2016} To highlight the two-magnon contribution in all samples the difference between $T = 30$ K and $310$ K has been filled with color. In Fig. \ref{Magnonfit} (b) this area (500 to 2000 cm$^{-1}$) has been plotted as a function of increased temperature, confirming that the spectral weight originates from magnetic scattering. 

The drastic change in lineshape of the two-magnon when entering the SRO phase can be qualitatively understood based on the quenching of LRO and the appearance of fluctuating pseudospins. REXS measurements on the same sample  \cite{Gretarsson_RIXS_LaDoping_2016} show a correlation length of only $\sim 8$ ${\rm\AA}$  at $T = 20$ K, this corresponds to approximately 3 sites correlation between pseudospins, less than what the broken-bond model requires for a two-magnon scattering process. As a consequence the two-magnon scattering loses intensity and becomes overdamped. 

\section{Discussion}

We start our discussion with a brief summary of our main experimental findings. Our Raman scattering study on La-doped \SIO\ revealed three significant features at low temperatures. Firstly, the lineshapes of the $A_{1g}$ mode at $\sim$188 cm$^{-1}$ and the $B_{2g}$ phonon mode at $\sim$191 cm$^{-1}$ remain strongly anisotropic in the doped samples. The low temperature renormalization of these lineshapes is suppressed alongside with the long-range magnetic order. Secondly, in parallel to this, the single magnon excitation is rapidly quenched, and only a strongly broadened two-magnon feature persists in the short-ranged ordered phase. Thirdly,  in the most highly doped sample the 395 cm$^{-1}$ $B_{2g}$ phonon mode exhibits a Fano line shape, an effect that was not seen in the undoped material.

Taken together, these experimental results indicate two different interaction channels for the phonons. The first channel, which dominates the coupling to the $\sim$188 cm$^{-1}$ $A_{1g}$ and the $\sim$191 cm$^{-1}$ $B_{2g}$ modes, arises from a continuum of pseudospin magnetic excitations. This continuum exists at high temperatures at all doping levels (in agreement with recent RIXS data~\cite{Gretarsson_RIXS_LaDoping_2016}). Once the long-range magnetic order sets in, the continuum condenses into a set of dispersive magnon modes, removing the anomalous self-energy of the corresponding phonon modes. In the x=0.10 sample, the gradual increase in $1/|q|$ with increasing temperature in both of those modes suggests that at higher temperatures the spectral weight of pseudospins fluctuations grows stronger.  This observation is in agreement with REXS measurements showing magnetic correlation lengths from $\sim$ 8 ${\rm\AA}$  at $T = 20$ K to $\sim$ 5 ${\rm\AA}$  at $T = 300$ K.\cite{Gretarsson_RIXS_LaDoping_2016} Furthermore, doping appears to have very little effect on the phonon lineshape at $T>T_N$ implying that the pseudospin degrees of freedom and their interactions with the lattice are doping-independent. The second channel, through which the 385 cm$^{-1}$ $B_{2g}$ mode acquires a Fano lineshape upon doping, appears on the other hand unrelated to magnetism and can thus be attributed to the doped carriers. Interestingly, a weak Fano asymmetry was seen for this stretching mode in the parent bilayer compound Sr$_3$Ir$_2$O$_7$~\cite{Gretarsson_TwoMagnon_2015}, but the origin of this lineshape remained unclear. The results we have presented here suggest that this bi-layer mode is also coupling to an electron-hole charge continuum. 

More generally our Raman measurements provide a glimpse into how the extra electrons, donated by the La atoms, alter the ground state of \SIO. Our measurements suggest that at higher energies ($\sim 50$ meV) doping induces a sizable electronic continuum, resulting in the Fano profile of the 395 cm$^{-1}$   $B_{2g}$ phonon modes. The presence of this new continuum has minimal impact on the 188 cm$^{-1}$ $A_{1g}$  and 191 cm$^{-1}$ $B_{2g}$ modes ($\sim 25$ meV), either due to a small spectral weight or a matrix element effect, suggesting that pseudospin-lattice coupling still dominates the renormalization of those phonons. This picture requires the presence of pseudospins in the $x=0.10$ sample, which is indeed supported by our two-magnon scattering results in Fig. \ref{Bimagnon}. Our observation of fluctuating 
pseudospins in the metallic x=0.10 sample, along with recent RIXS results,\cite{Gretarsson_RIXS_LaDoping_2016} raises the interesting possibility that magnetic correlations play an important role in the formation of the observed pseudogap in metallic \SLIO\ sample.\cite{Baumberger_PRL_2015} Our results are also compatible with an inhomogeneous phase-separated state suggested by recent STM experiment,\cite{MPAllan_arXiv_2016} where more metallic regions are nucleated around clusters of dopant atoms while insulating regions are found further away.

We end our discussion by looking at the strong temperature and doping dependence of some of the other $A_{1g}$ phonon modes. In Fig. \ref{Allphonon} (a) the two weak modes at 337 cm$^{-1}$ and 342 cm$^{-1}$ broaden rapidly with increasing temperature. Upon 1.5$\%$ La-doping these two modes have already emerged with the higher energy side of the 278 cm$^{-1}$ $A_{1g}$  mode and in the $x=0.10$ sample they are undetectable (see Fig. \ref{A1g_newPh}) . The rapid doping-dependence of these two modes was also observed in Sr$_2$IrO$_{4-\delta}$ samples,\cite{Nakheon_Growth_2016} suggesting that they are intimately connected to the magnetic order. It seems however unlikely that these modes are a results of folding of phonon modes elsewhere in the Brillouin Zone since the magnetic order in \SIO\ is not accompanied by structural changes. \cite{Sr214_Structure} We note that a double peak structure has been observed in optical conductivity measurements  (324 and 339 cm$^{-1}$) which only develops at low temperatures \cite{Propper_Optics:PRB2016,Moon_Optics:PRB2009} but the origin of those modes has not been discussed. 

Looking at the 278 cm$^{-1}$ mode, it is clear that its broad and unusual lineshape is also closely related to the loss of long range magnetic order. Indeed this mode shows very little doping dependence at $T = 340$ K, a temperature which is well above $T_N$, while being quite different at low temperatures. The exact origin of the strong dependence in these three modes (278 , 337, and 342 cm$^{-1}$) is unclear. We do however point out that the 278 cm$^{-1}$ $A_{1g}$  mode modulates the in-plane $\rm{Ir-O-Ir}$ bond angle and could therefore be more susceptible to magnetic order given the link between the in-plane canted moments and the rotation of the IrO$_6$ octahedra.\cite{Calder_Elastic:PRB2015} 
In light of the observed correlations between these phonons and magnetism, the reported splitting of the 278 cm$^{-1}$ A$_{1g}$  mode in Sr$_2$Ir$_{1-x}$Ru$_x$O$_4$ \cite{Sr214_Ru_Raman} might be understood based on the recent observation of two different magnetic structures in Sr$_2$Ir$_{1-x}$Ru$_x$O$_4$, with $ab$-plane and $c$-axis aligned \jeff=1/2 pseudospin moments, respectively.\cite{Calder_Elastic:PRB2015} We point out that a similar effect was seen in Sr$_2$IrO$_{4-\delta}$ samples \cite{Nakheon_Growth_2016}, underscoring that magnetism, and not the different Ir(Ru)-O-Ir(Ru) angle, is the driver behind this behavior.

\section{Conclusion}

To summarize, La doping rapidly obliterates the single-magnon feature and greatly weakens two-magnon excitations in the Raman spectra of  \SLIO\, along with the disappearance of magnetic long-range order. However, the lineshapes of low-energy phonons reveal persistent pseudospin-lattice coupling in the metallic state, while the continuum of charge excitations in the metallic state gives rise to additional lineshape anomalies of a higher-energy phonon. Our Raman scattering results are consistent with recent RIXS data that indicate well-defined pseudospin excitations in  \SLIO. \cite{Gretarsson_RIXS_LaDoping_2016,Dean_arXiv_2016} They thus underscore the persistence of strong electronic correlations in metallic iridates, in close analogy to the doped cuprates. \cite{LeTacon_NatPhys_2011, MPDean_NatMat_2013,Minola_PRL_2015}

\acknowledgements{N. H. Sung was supported by the Alexander von
Humboldt Foundation. We acknowledge financial support by the DFG under grant TRR80.}

\end{document}